\documentclass[%
 aip,
groupedaddress,
 amsmath,amssymb,
 reprint,%
]{revtex4-1}

\usepackage{graphicx}
\usepackage{dcolumn}
\usepackage{bm}

\usepackage{xparse, physics, xcolor}

\usepackage[utf8]{inputenc}
\usepackage[T1]{fontenc}
\usepackage{mathptmx}
\usepackage{etoolbox}

\newcommand{\ra}{{\rm Ra}}
\newcommand{\pe}{{\rm Pe}}
\newcommand{\pec}{{\mathcal{P}}}
\newcommand{\equa}[1]{equation~(\ref{#1})}

\newcommand{\equas}[1]{equations~(\ref{#1})}
\newcommand{\equass}[2]{equations~(\ref{#1})--(\ref{#2})}
\newcommand{\equasa}[2]{equations~(\ref{#1}){ }and{ }(\ref{#2})}
\newcommand{\Equa}[1]{Equation~(\ref{#1})}

\newcommand{\Equas}[1]{Equations~(\ref{#1})}

\newcommand{\Equasa}[2]{Equations~(\ref{#1}){ }and{ }(\ref{#2})}
\newcommand{\eqn}[2]{\begin{gather}
\begin{aligned}
#1
\label{#2}
\end{aligned}
\end{gather}
}

\newcommand{\spl}[2]{\begin{gather}
#1
\label{#2}
\end{gather}
}

\newcommand{\gat}[2]{\begin{subequations}\label{#2}\begin{gather}
#1
\end{gather}\end{subequations}
}
\makeatletter
\def\@email#1#2{%
 \endgroup
 \patchcmd{\titleblock@produce}
  {\frontmatter@RRAPformat}
  {\frontmatter@RRAPformat{\produce@RRAP{#1\href{mailto:#2}{#2}}}\frontmatter@RRAPformat}
  {}{}
}%
\makeatother

\begin{document}


\title[Space--Evolving Instability]{Time--Evolving to Space--Evolving Thermal Instability\\ of a Porous Medium Flow}
\author{A. Barletta}
\affiliation{Department of Industrial Engineering, Alma Mater Studiorum Universit\`a di Bologna,\\ Viale Risorgimento 2, 40136 Bologna, Italy}%
 \homepage{https://www.unibo.it/sitoweb/antonio.barletta/en}
 \email{antonio.barletta@unibo.it}

\date{\today}

\begin{abstract}
The Rayleigh--B\'enard instability of the stationary throughflow in a horizontal porous layer, also known as Prats' problem, is here analysed in a fresh new perspective. In fact, the classical analysis of the linear instability, carried out by employing time--evolving and space--periodic Fourier modes, is here reconsidered by focussing on the effects of time--periodic and space--evolving modes. The basic stationary flow is assumed to be perturbed by a localised source of perturbation which is steady--periodic in time. Then, the spatial development of such perturbations is monitored in order to detect their possible amplification or decay in their direction of propagation. Accordingly, the spatial stability/instability threshold is determined. The study is carried out by employing a Fourier transform formalism, where the transformed variable is time.
\end{abstract}

\maketitle

\section{Introduction}
The linear stability analysis of shear flows can be carried out either by employing space--periodic perturbation modes with a real wavenumber and a complex angular frequency or by utilising time--periodic modes with a complex wavenumber and a real angular frequency \cite{schmid2012stability}. The former approach, also called temporal stability analysis, is the classical linear stability analysis aimed to determine the time--growth rate of the perturbation modes, the neutral stability curve for the onset of convective instability and the critical parameters for the existence of unstable modes. On the other hand, the latter type of analysis leads to the evaluation of the spatial stability/instability, namely the existence of spatially decaying/growing modes in their direction of propagation.

There exists a wide literature about the spatial stability analysis of jets and mixing layers \citep{betchov1966spatial, keller1973spatial, ortiz2002spatial, afzaal2015temporal}. The importance of these studies arises mainly from their applications in aerodynamics and in the dynamics of oceanic streams. The development of spatial modes of instability in internal flows with permeable boundaries is discussed by \citet{casalis1998spatial}, and by \citet{griffond2000spatial}. In these papers, the analysis of spatial instability focusses on basic flow conditions where a plane channel or circular duct features permeable boundaries with fluid injection. \citet{gill1965behaviour}, as well as \citet{garg_rouleau_1972}, examines the spatial stability of the Hagen--Poiseuille flow in a circular pipe. \citet{garg_rouleau_1972} demonstrate that this basic flow is spatially stable according to the linear analysis to all Reynolds number (at least, up to $10000$), thus confirming the well--known result of the stability analysis to real wavenumber and complex angular frequency modes. {Other closely related studies were published recently, where the concept of spatially--developing time--periodic modes is further investigated \cite{sircar2019spatiotemporal, colanera2021modal, hasan2021numerical}.}

As pointed out by \citet{gill1965behaviour}, the importance of a type of linear analysis involving ``a disturbance of a fixed frequency that decays with downstream distance'', {\em i.e.} of the spatial stability analysis, is justified as ``in experiments in which disturbances are introduced into the flow in a controlled manner, a constant frequency--generating device is commonly used''. The same concept is emphasised in chapter~7 of \citet{schmid2012stability} when the authors say that the ``excitation of spatially growing disturbances by a vibrating ribbon is
a common experimental technique to investigate the response of a shear layer to harmonic forcing''. The clue of the spatial approach to the linear stability analysis is that the flow instability is not necessarily an application of Lyapunov's definition, where instability means that an arbitrarily small perturbation of the initial state at time $t=0$ yields a perturbation amplitude growing in time. The spatial stability analysis interchanges the roles of space and time, inasmuch as a perturbed initial condition at $t=0$ is replaced by a perturbed condition set at a given spatial position, say $x=0$. This given spatial position is the place where the ``constant frequency--generating device'' mentioned by \citet{gill1965behaviour}, or the ``vibrating ribbon'' considered by \citet{schmid2012stability}, is localised. Such a device is the pointlike source of harmonic forcing and, as a consequence, the cause of the perturbation. 

Many authors have also highlighted the higher level of mathematical complexity typical of the spatial stability analysis, with respect to the classical stability analysis of space--periodic and time--growing modes. This reflection stems from the intrinsic characteristics of flow systems where the governing equations are, usually, second--order in space and first--order in time. There are obvious exceptions to this rule when it comes, say, to viscoelastic flows or acoustic wave dynamics. In fact, a higher differential order in a coordinate means a higher degree of the wavenumber in the dispersion relation, and likewise for the differential order in time and the degree of the angular frequency. The temporal stability analysis aims to provide an explicit determination of the (complex) angular frequency, while the spatial stability analysis aims to obtain an explicit evaluation of the (complex) wavenumber. The larger is the degree of the variable to be solved for in the dispersion relation, the larger is the number of solution branches to be accounted for in the stability analysis. This extra mathematical difficulty in the  spatial stability analysis is pointed out, for instance, in \citet{schmid2012stability} where the authors say: ``the problem of determining the spatial stability is given by an eigenvalue problem where the eigenvalue appears nonlinearly''.

The spatial stability analysis of fluid flows in saturated porous media is a relatively new area where only a few explorations are available in the literature to date. The most significant bulk of papers where a complex wavenumber is employed in a stability analysis regards the absolute instability \citep{delache2007spatio, Brevdo2009, DiazBrevdo2011, barletta2017absolute, barletta2017convective, barletta2019routes, de2019identifying, schuabb2020two}. However, the focus of the transition to absolute instability is on the dynamics of wavepackets mathematically modelled as Fourier integrals that linearly combine time--growing modes \citep{barletta2019routes}. In this framework, the use of a complex wavenumber is only a mathematical tool requested by the steepest--descent approximation of the Fourier integrals at large times \citep{barletta2019routes}.
As such, the absolute instability analysis is nothing but a step within the temporal stability analysis entirely embodying Lyapunov's concept. In other words, envisaging a complex wavenumber does not imply necessarily carrying out a spatial stability analysis. \citet{tyvand2020laterally} have recently studied the characteristics of spatial modes having a complex wavenumber as a tool to lay out an asymptotic analysis of the spatial decay of perturbations in the domains
of lateral penetration within a horizontal porous layer. In fact, these authors aim to study, for a porous layer, a model marginal state of convection which is localised in space \citep{tyvand2020laterally}.

The objective of the investigation presented in the forthcoming sections is to provide a spatial analysis of the Rayleigh--B\'enard instability in a horizontal porous layer subjected to a basic horizontal throughflow. Such an instability is also known as Prats' problem \citep{Prats}, from the name of the author that developed the temporal stability analysis for this basic configuration. The spatial stability of Prats' problem is not available in the literature although a special case, {\em i.e.} that where no throughflow is present in the basic state, is discussed in a recent paper \citep{barletta2021spatially}. Since throughflow in Prats' problem is parametrised by the P\'eclet number, $\pe$, the special case examined by \citet{barletta2021spatially} corresponds to the limit $\pe \to 0$. The aim is to extend this investigation by allowing for a general $\pe \ne 0$ and for a three-dimensional stability analysis, as that carried out by \citet{barletta2021spatially} is two--dimensional. 
As shown in Prats' paper \citep{Prats}, the temporal analysis of the convective instability for Prats' problem discloses no really new features as compared to the analysis of the Rayleigh--B\'enard instability in a horizontal porous medium. The reason is that the basic throughflow, being uniform, just affects the phase velocity of the perturbing modes, while the neutral stability condition and the critical values of the Rayleigh number and wavenumber at instability onset do not depend on the flow rate. Things get different when it comes to the analysis of absolute instability \citep{barletta2019routes}. Just the same nontrivial role played by the basic throughflow emerges also for the spatial instability, as it will become clear in the forthcoming sections.
Since the method involving spatial modes could be unfamiliar for readers interested in convection processes and buoyant flows in porous media, the presentation of the study is carried out step--by--step by briefly recalling the main features of the temporal stability analysis and introducing the spatial stability analysis of Prats' problem. The motivation of this approach is also the need of justifying, mathematically, why the dual methods with either a real wavenumber and a complex frequency or a complex wavenumber and a real frequency are both conceivable. The mathematical tool leading to such schemes is the Fourier transform involving either a coordinate or time in the two approaches. Stressing this point possibly fills a gap in some pedagogical treatments of this subject available in the literature.
The use of space--evolving modes leads to the conclusion that such modes can yield a destabilisation of the basic throughflow whatever small is the positive value of the Rayleigh number, $\ra$, which parametrises the temperature difference between the boundaries and, thus, the heating--from--below mechanism causing the Rayleigh--B\'enard instability. As expected, for a negative $\ra$ (heating from above) spatial instability turns out to be impossible.

\section{Mathematical Model}

Let us consider a horizontal porous channel bounded by the two planes $y=0$ and $y=H$, where the $y$ axis is oriented vertically. 
The two boundary planes $y=0$ and $y=H$ are impermeable and isothermal with temperatures $T_h$ and $T_c < T_h$, respectively, with $c$ and $h$ standing for cold and hot. In the unperturbed state, the saturating fluid flows with a uniform and constant velocity $U_0$. 

The Oberbeck--Boussinesq approximation is employed within a model of seepage flow based on Darcy's law. 
Thus, the dimensionless governing equations are written as
\gat{
\pdv{u}{x} + \pdv{v}{y} + \pdv{w}{z} = 0,\label{1a}
\\
\pdv{w}{y} - \pdv{v}{z} = - \ra \, \pdv{T}{z} ,\label{1b}
\\
\pdv{u}{z} - \pdv{w}{x} = 0 ,\label{1c}
\\
\pdv{v}{x} - \pdv{u}{y} = \ra \, \pdv{T}{x} ,\label{1d}
\\
\pdv{T}{t} + u\, \pdv{T}{x} + v\, \pdv{T}{y} + w\, \pdv{T}{z} = \pdv[2]{T}{x} + \pdv[2]{T}{y} + \pdv[2]{T}{z} .\label{1e}
}{1}
\Equas{1} express the local balances of mass, momentum and energy, respectively. In particular, \equass{1b}{1d} are the vorticity formulation of the momentum balance. The velocity components in the $(x,y,z)$ directions are $(u,v,w)$, $t$ is the time, while $T$ is the temperature. The porous medium definition of Rayleigh number is given by
\eqn{
\ra = \frac{g\, \beta \, \qty(T_h - T_c) \, K\, H}{\nu\, \alpha},
}{2}
where $g$ is the modulus of the gravitational acceleration, $\beta$ is the coefficient of fluid thermal expansion, $K$ is the permeability, $\nu$ is the kinematic viscosity and $\alpha$ is the average thermal diffusivity of the saturated porous medium. 

The dimensionless formulation expressed through \equas{1} is obtained by scaling the dimensional quantities as
\spl{
\frac{\qty(x,y,z)}{H} \to \qty(x,y,z) \qc \frac{t}{\sigma H^2/\alpha} \to t ,
\nonumber\\
\frac{\qty(u,v,w)}{\alpha/H} \to \qty(u,v,w) \qc \frac{T - T_c}{T_h - T_c} \to T .
}{3}
The dimensionless parameter $\sigma$ is the ratio between the volumetric heat capacity of the saturated porous medium and the volumetric heat capacity of the fluid. Hereafter, all fields, coordinates and time will be intended as dimensionless through such a scaling.

\subsection{The boundary conditions and the basic flow}
The boundaries $y=0$ and $y=1$ are subject to the same thermal and dynamic conditions employed in the paper by \citet{Prats}, namely
\spl{
v = 0 \qc T = 1 \qif y = 0,
\nonumber\\
v = 0 \qc T = 0 \qif y = 1.
}{4}
A stationary flow solution of \equasa{1}{4} is allowed with the basic flow velocity lying on the horizontal $xz$ plane and inclined an angle $\phi$ to the $x$ axis, namely 
\eqn{
u_b = \pe \cos\phi \qc v_b = 0 \qc w_b = \pe \sin\phi  \qc T_b = 1 - y .
}{5}
The basic flow fields are identified with the subscript $b$ and depend on the P\'eclet number,
\eqn{
\pe = \frac{U_0 \, H}{\alpha} ,
}{6}
as well as on the angle $\phi$.

\subsection{Dynamics of the small--amplitude perturbations}
The basic flow (\ref{5}) is assumed to be weakly perturbed as,
\spl{
u = \pe\,\cos\phi + \epsilon \, U \qc v = \epsilon \, V \
\nonumber\\
w = \pe\,\sin\phi + \epsilon \, W \qc T = 1 - y + \epsilon \, \theta ,
}{7}
where the perturbations $(U,V,W,\theta)$ are modulated by a small parameter $\epsilon$, such that the substitution of \equa{7} in \equasa{1}{4} yields the linearisation,
\gat{
\pdv{U}{x} + \pdv{V}{y} + \pdv{W}{z} = 0,\label{8a}
\\
\pdv{W}{y} - \pdv{V}{z} = - \ra \, \pdv{\theta}{z} ,\label{8b}
\\
\pdv{U}{z} - \pdv{W}{x} = 0 ,\label{8c}
\\
\pdv{V}{x} - \pdv{U}{y} = \ra \, \pdv{\theta}{x} ,\label{8d}
\\
\pdv{\theta}{t} + \pe\,\cos\phi\, \pdv{\theta}{x}  + \pe\,\sin\phi\, \pdv{\theta}{z} - V 
\nonumber\\
\displaybreak
= \pdv[2]{\theta}{x} + \pdv[2]{\theta}{y} + \pdv[2]{\theta}{z} ,\label{8e}
\\
V = 0 \qc \theta = 0 \qif y = 0, 1.\label{8f}
}{8}
In \equas{8}, terms of order $\epsilon^2$ were neglected, so that $\epsilon$ could be simplified in all the governing equations. 

The linear dynamics of the perturbations can be studied by assuming their propagation in any horizontal direction. Such an arbitrary horizontal direction can be chosen as the $x$ axis without any loss of generality, provided that the angle $\phi$ between the basic velocity and the $x$ axis is considered arbitrary. Then, \equas{8} can be turned into a two--dimensional formulation where the perturbations are assumed to be independent of $z$, {\em i.e.} the axis perpendicular to the propagation direction of the perturbations.
As a consequence, the velocity perturbations can be expressed in terms of a streamfunction, $\Psi$, namely
\eqn{
U = \pdv{\Psi}{y} \qc V = - \pdv{\Psi}{x} .
}{9}
Then, a reformulation of \equas{8} is obtained
\gat{
\pdv[2]{\Psi}{x} + \pdv[2]{\Psi}{y} + \ra \, \pdv{\theta}{x} = 0 ,\label{10a}
\\
\pdv[2]{\theta}{x} + \pdv[2]{\theta}{y} - \pdv{\theta}{t} - \pec\, \pdv{\theta}{x} - \pdv{\Psi}{x} = 0 ,
\\
\Psi = 0 \qc \theta = 0 \qif y = 0, 1,\label{10c}
}{10}
where $\pec = \pe\,\cos\phi$ is a modified P\'eclet number which depends on the basic flow rate and on its inclination to the $x$ axis.

{It is to be emphasised that having turned the governing equations for the perturbing modes to a two--dimensional form does not alter, from the physical viewpoint, the three--dimensional nature of the stability analysis. In fact, by changing the angle $\phi$, one tests the effect of a generic plane wave propagating along any inclined direction relative to the basic flow direction. Furthermore, assuming a $z$ independence just characterises such perturbation modes as plane waves, but does not constrain in any manner their propagation direction.}

We now express $\Psi$ and $\theta$ in terms of Fourier series as
\spl{
\Psi(x,y,t) = \sum_{n=1}^{\infty} \Psi_n(x,t) \, \sin(n \pi y),
\nonumber\\
\theta(x,y,t) = \sum_{n=1}^{\infty} \theta_n(x,t) \, \sin(n \pi y),
}{11}
where $\Psi_n$ and $\theta_n$ can be evaluated as
\spl{
\Psi_n(x,t) = 2\int_{0}^{1} \Psi(x,y,t) \, \sin(n \pi y)\, \dd y,
\nonumber\\
\theta_n(x,t) = 2\int_{0}^{1} \theta(x,y,t) \, \sin(n \pi y)\, \dd y.
}{12}
Thus, \equa{10c} is satisfied and we can reformulate \equas{10} as
\gat{
\pdv[2]{\Psi_n}{x} - \qty(n \pi)^2 \Psi_n + \ra \, \pdv{\theta_n}{x} = 0 ,
\\
\pdv[2]{\theta_n}{x} - \qty(n \pi)^2 \theta_n - \pdv{\theta_n}{t} - \pec\, \pdv{\theta_n}{x} - \pdv{\Psi_n}{x} = 0 .
}{13}

\section{Time--evolving Fourier modes}\label{timdev}
A possible approach to the solution of \equas{13} is based on the Fourier transform, with the $x$ coordinate being the transformed variable. We define
\spl{
\tilde\Psi_n(k,t) = \frac{1}{\sqrt{2\pi}} \int_{-\infty}^{\infty} \Psi_n(x,t) \, e^{-i k x}\, \dd x,
\nonumber\\
\tilde\theta_n(k,t) = \frac{1}{\sqrt{2\pi}} \int_{-\infty}^{\infty} \theta_n(x,t) \, e^{-i k x}\, \dd x,
}{14}
so that the inverse transform yields
\spl{
\Psi_n(x,t) = \frac{1}{\sqrt{2\pi}} \int_{-\infty}^{\infty} \tilde\Psi_n(k,t) \, e^{i k x}\, \dd k,
\nonumber\\
\theta_n(x,t) = \frac{1}{\sqrt{2\pi}} \int_{-\infty}^{\infty} \tilde\theta_n(k,t) \, e^{i k x}\, \dd k.
}{15}
By utilising the properties of Fourier transforms, we can rewrite \equas{13} as
\spl{
\pdv{\tilde\theta_n}{t} = - \qty[k^2 + \qty(n \pi)^2 - \frac{k^2\, \ra}{k^2 + \qty(n \pi)^2} + i k\, \pec] \tilde\theta_n , 
\nonumber\\
\text{with} \quad \tilde\Psi_n = \frac{i k\, \ra}{k^2 + \qty(n \pi)^2} \, \tilde\theta_n .
}{16}
The solution for $\tilde\theta_n$ is a simple exponential in time,
\spl{
\tilde\theta_n(k,t) = \tilde\theta_n(k,0) \, e^{\lambda(k) t} , 
\nonumber\\
\text{with} \quad \lambda(k) = - k^2 - \qty(n \pi)^2 + \frac{k^2\, \ra}{k^2 + \qty(n \pi)^2} - i k\, \pec .
}{17}
If we use \equas{11}, (\ref{12}) and (\ref{14})--(\ref{17}), we can evaluate the time evolution of a perturbation assigned at time $t=0$. The steps are as follows:
\begin{enumerate}
\item Start from an initial perturbation $\theta(x,y,0)$;
\item Evaluate $\theta_n(x,0)$ by employing \equa{12};
\item Evaluate $\tilde\theta_n(k,0)$ by employing \equa{14};
\item Evaluate $\tilde\theta_n(k,t)$ by employing \equa{17};
\item Evaluate $\tilde\Psi_n(k,t)$ by employing \equa{16};
\item Evaluate $\Psi_n(x,t)$ and $\theta_n(x,t)$ by employing \equa{15};
\item Evaluate $\Psi(x,y,t)$ and $\theta(x,y,t)$ by employing \equa{11}.
\end{enumerate}
Such seven steps highlight that the initial perturbation, fixed arbitrarily, is a temperature perturbation. One cannot fix independently also an initial perturbation for the streamfunction. The reason is that $\Psi$ and $\theta$ are not independent inasmuch as they are governed by \equa{10a}.
This seven--steps method embodies Lyapunov's concept of instability for an equilibrium state. When it comes to fluid flow, an equilibrium state is a stationary solution of the governing equations. A small amplitude perturbation at time $t=0$ may either yield an amplification in time or a damping in time or even be time--independent. Lyapunov's instability is the former case, while stability is the second or third case. The concept is simple and well--known: modify, but only slightly, the initial condition of the flow system and see what happens. 
At this point, the way one monitors the evolution in time of an initially prescribed perturbation makes the difference between the convective instability and what is called absolute instability.

\subsection{Convective instability}\label{coninst}
Imagine an initial perturbation set at time $t=0$ with the shape of a monochromatic plane wave, {\em i.e.} a plane wave with a given wavenumber. Strictly speaking this means a function that is not absolutely--integrable over the real $x$ axis. Thus, its Fourier transform does not exist if not in terms of a generalised function, or distribution. In fact, the Fourier transform of a plane wave is a Dirac's delta function, {\em i.e.} something proportional to $\delta\qty(k - k_p)$, where $k_p$ is the wavenumber of the plane wave. If one aims to monitor the time evolution of a plane wave with a given wavenumber $k$, then one is seeking the convective instability of the flow system. The ultimate evolution of an initial condition $\theta(x,y,0)$ expressed through a plane wave $\exp\!\qty(i k x)$ is easily judged by employing \equa{17}, with $\Re(\lambda) = \gamma$ and $\Im(\lambda)= -\omega$. Then, one has
\eqn{
\gamma = - k^2 - \qty(n \pi)^2 + \frac{k^2\, \ra}{k^2 + \qty(n \pi)^2} \qand \omega = k\, \pec .
}{18}
With a direct physical meaning for both $\gamma$ and $\omega$: the former is the exponential growth rate, the latter is the angular frequency of the time--evolving wave. Convective instability is when $\gamma >0$ or, equivalently,
\eqn{
\ra > \frac{\qty[k^2 + \qty(n \pi)^2]^2}{k^2} .
}{19}
The lowest branch of instability described by \equa{19} is with $n=1$. Thus, the onset of the convective instability occurs by the modes $n=1$, at the threshold
\eqn{
\ra = \frac{\qty(k^2 + \pi^2)^2}{k^2} ,
}{20}
also known as neutral stability condition. The least value of $\ra$ at neutral stability is the minimum, or critical, value $\ra_c = 4\pi^2$ occurring when $k = k_c = \pi$. Such results are all well--known starting from the paper by \citet{Prats} and are reported in detail, say, in the books by \citet{Straughan}, by \citet{NieldBejan2017} and by \citet{barletta2019routes}. An evident feature of the convective instability for Prats' problem is that its onset is a condition independent of the  modified P\'eclet number, $\pec$, as it is evident from \equasa{19}{20}. This is not true anymore if we are to define the transition to absolute instability.

\subsection{Absolute instability}\label{absinstity}
If the initial condition is not a wavelike signal, $\theta(x,y,0)$, but a function of $x$ absolutely integrable over the real $x$ axis, then we are monitoring the time development of a localised wavepacket. In this case, we do not need the theory of distributions as the Fourier transform is defined in the usual sense, {\em viz.} in terms of functions.

In fact, the absolute instability occurs when
\eqn{
\lim_{t \to +\infty} \qty|\Psi(x,y,t)| = \infty \qand \lim_{t \to +\infty} \qty|\theta(x,y,t)| = \infty .
}{21}
Thus, on account of \equa{11}, we are to establish the large--$t$ behaviour of the Fourier integrals on the right hand side of \equa{15} which, by employing \equasa{16}{17}, can be rewritten as
\spl{
\Psi_n(x,t) = \frac{1}{\sqrt{2\pi}} \int_{-\infty}^{\infty} \tilde\Psi_n(k,0) \, e^{\lambda(k) t} \, e^{i k x}\, \dd k,
\nonumber\\
\theta_n(x,t) = \frac{1}{\sqrt{2\pi}} \int_{-\infty}^{\infty} \tilde\theta_n(k,0) \, e^{\lambda(k) t} \, e^{i k x}\, \dd k.
}{22}
As widely discussed by several authors (see, for instance, \citet{barletta2019routes},  \citet{lingwood1997application} or \citet{suslov2006numerical}), stationary phase mathematical techniques such as the steepest--descent approximation \cite{ablowitz2003complex} can be efficiently utilised in the evaluation of the limiting behaviour for $t \to +\infty$ of the integrals employed in \equa{22}. In this way, the condition of absolute instability is identified when the real part of $\lambda(k_0)$ is positive. Here, $k_0$ is a saddle point of $\lambda(k)$, that is a zero of its derivative \citep{lingwood1997application, suslov2006numerical, barletta2019routes}, namely $\dd\lambda(k)/\dd k$. We do not go any further into the details of the absolute instability analysis for the Prats' problem and we refer the reader to the detailed studies available in the literature \cite{delache2007spatio, barletta2019routes}. We just say that the transition to an absolutely unstable regime occurs when the Rayleigh number $\ra$ exceeds a threshold value, $\ra_a$, coincident with the critical value $\ra_c$ for $\pec=0$. When $\pec>0$, one has $\ra_a >\ra_c$ with $\ra_a$ monotonically increasing with $\pec$.

\section{Space--evolving Fourier modes}\label{spaevo}
An alternative solution method for \equas{13} is again based on the Fourier transform, but with the time $t$ as the transformed variable. Thus, we write
\spl{
\hat\Psi_n(x,\omega) = \frac{1}{\sqrt{2\pi}} \int_{-\infty}^{\infty} \Psi_n(x,t) \, e^{i \omega t}\, \dd t,
\nonumber\\
\hat\theta_n(x,\omega) = \frac{1}{\sqrt{2\pi}} \int_{-\infty}^{\infty} \theta_n(x,t) \, e^{i \omega t}\, \dd t,
}{23}
so that the inverse transform yields
\spl{
\Psi_n(x,t) = \frac{1}{\sqrt{2\pi}} \int_{-\infty}^{\infty} \hat\Psi_n(x,\omega) \, e^{-i \omega t}\, \dd \omega,
\nonumber\\
\theta_n(x,t) = \frac{1}{\sqrt{2\pi}} \int_{-\infty}^{\infty} \hat\theta_n(x,\omega) \, e^{-i \omega t}\, \dd \omega.
}{24}
On account of the properties of Fourier transforms, we can rewrite \equas{13} as%
\gat{
\pdv[2]{\hat\Psi_n}{x} - \qty(n \pi)^2 \hat\Psi_n + \ra \, \pdv{\hat\theta_n}{x} = 0 ,
\\
\pdv[2]{\hat\theta_n}{x} - \qty(n \pi)^2 \hat\theta_n + i \omega \, \hat\theta_n - \pec\, \pdv{\hat\theta_n}{x} - \pdv{\hat\Psi_n}{x} = 0 .
}{25}
The solution for $\hat\Psi_n$ and $\hat\theta_n$ is given by exponentials in $x$,
\spl{
\hat\Psi_n(x,\omega) = \hat\Psi_n(0,\omega) \, e^{\eta(\omega) x} ,
\nonumber\\
\hat\theta_n(x,\omega) = \hat\theta_n(0,\omega) \, e^{\eta(\omega) x} ,
}{26}
with
\eqn{
\hat\Psi_n = - \frac{\eta\, \ra}{\eta^2 - \qty(n \pi)^2} \, \hat\theta_n ,
}{27}
provided that $\eta(\omega)$ satisfies the relation
\eqn{
\eta^2 - \qty(n \pi)^2 + i \omega + \frac{\eta^2\, \ra}{\eta^2 - \qty(n \pi)^2} - \eta\, \pec = 0 .
}{28}
The first apparent feature of \equa{28} is that it defines function $\eta(\omega)$ only implicitly. Turning such an implicit definition into an explicit expression would give rise to four solution branches. The second feature regards the use of these solutions.

By employing \equas{11}, (\ref{12}), (\ref{23}), (\ref{24}) and (\ref{26})--(\ref{28}), one can evaluate the space development of a perturbation which is assigned at the initial cross--section, $x=0$. The steps are as follows:
\begin{enumerate}
\item Start from an initial perturbation $\theta(0,y,t)$;
\item Evaluate $\theta_n(0,t)$ by employing \equa{12};
\item Evaluate $\hat\theta_n(0,\omega)$ by employing \equa{23};
\item Evaluate $\hat\theta_n(x,\omega)$ by employing \equasa{26}{28};
\item Evaluate $\hat\Psi_n(x,\omega)$ by employing \equa{27};
\item Evaluate $\Psi_n(x,t)$ and $\theta_n(x,t)$ by employing \equa{24};
\item Evaluate $\Psi(x,y,t)$ and $\theta(x,y,t)$ by employing \equa{11}.
\end{enumerate}

The main difference with respect to Section~\ref{timdev} is that, here, we are monitoring the space development of a perturbation acting at $x=0$. Then, we are setting up a concept sharply different from the Lyapunov approach to instability. The latter approach is based on the sensitivity to the initial condition at $t=0$, while we are now assessing the sensitivity to the initial condition at $x=0$. 

{\Equa{23} makes it evident that time is now assumed to range from $-\infty$ to $+\infty$. Such a scheme is typical of steady--periodic phenomena where there is no starting time or initial time, so that $t=0$ has no special physical meaning. This is obviously different from the type of analysis developed in Section~\ref{timdev} where the system reaction to an initially imposed perturbation, set at $t=0$, is tested.}

An important comment is relative to step 4 listed above. In fact, $\hat\theta_n(x,\omega)$ can be evaluated only if $\eta(\omega)$ were uniquely defined by solving \equa{28}. However, this is not the case since \equa{28} admits four roots. The point is that the initial condition given by the assignment of $\theta(0,y,t)$ is not sufficient. Further initial conditions are needed, such as the specifications of $\Psi(0,y,t)$ and of the derivatives of $\theta$ and $\Psi$ with respect to $x$ evaluated at $x=0$. Anyway, the fourfold nature of the initial condition to be prescribed at $x=0$ does not alter the core of the reasoning given above: with the Fourier transform formulation defined by \equasa{23}{24}, we are testing the sensitivity to the entrance condition for the perturbations set at $x=0$. The initial condition at $x=0$ induces an evolution in space both in the negative $x$ direction and in the positive $x$ direction. This is a marked difference between time evolution, subject to the causality principle, and space evolution where there is no preferred direction and the physical domain is $-\infty < x < + \infty$.

Perturbations originated at $x=0$ and propagating rightward $(x>0)$ or leftward $(x<0)$ can be wavepackets made by many single--frequency modes, where the superposition is mathematically given by the Fourier integrals in \equa{24}. Alternatively, one can study the dynamics of each single--frequency mode, labelled by the value of $\omega$, evolving from $x=0$ either in the rightward $x$ direction or in the leftward $x$ direction. The latter path will be that pursued in the forthcoming study.
It might be emphasised that the actual cause of the single--frequency perturbations is a harmonic (time--periodic) source placed at $x=0$. The direction of propagation is determined by the phase velocity, $\omega/k$. Thus, if a perturbation is characterised by $\omega/k>0$ (wave propagating in the positive $x$ direction), its origin is in the source position $x=0$ and its domain of existence is the positive $x$ axis. On the contrary, if a perturbation is characterised by $\omega/k<0$ (wave propagating in the negative $x$ direction), its origin is again in the source position $x=0$ and its domain of existence is the negative $x$ axis. 
The toy model proposed in Appendix~\ref{App} can be helpful for exemplifying this point.

\begin{figure*}
\centering
\includegraphics[width=0.8\textwidth]{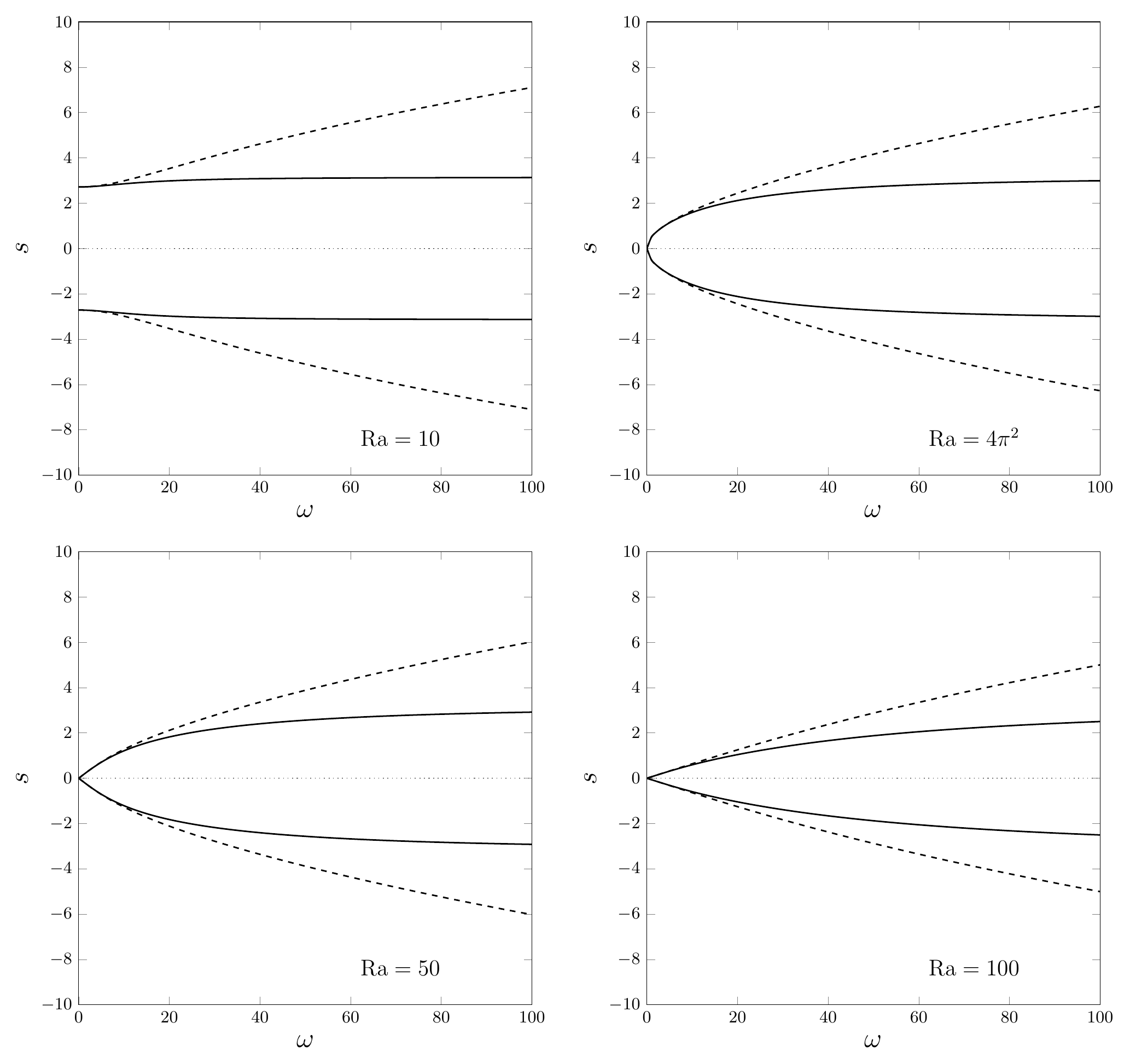}
\caption{\label{fig1}Plots of $s$ versus $\omega$ for $\pec=0$ and different values of $\ra$. Dashed lines denote spatially stable branches $(s\,k < 0)$, while solid lines denote spatially unstable branches $(s\,k > 0)$.}
\end{figure*} 

\subsection{Spatial instability}
We have pointed out that the focus is on testing the system reaction to a time--dependent perturbation acting at the entrance cross--section $x=0$. The absolute value of the perturbation has to be integrable over the real $t$ axis, which means a perturbation localised in time. However, this restrictive characterisation can be relaxed should one extend the Fourier transform to the domain of distributions. Then, a sinusoidal signal is allowed. In this case, $\hat\theta_n(0,\omega)$ is proportional to a Dirac's delta, $\delta\qty(\omega - \omega_p)$, where $\omega_p$ is the angular frequency of the wavelike perturbation, {\em i.e.} of the sinusoidal signal. In analogy with the analysis performed in Section~\ref{coninst}, the growth in space of the periodic sinusoidal signal acting at $x=0$ and having an angular frequency $\omega$ develops with a growing amplitude pattern for $x>0$ when $\Re\!\qty[\eta(\omega)] > 0$. On the other hand, for $x>0$, the perturbation is damped in space if $\Re\!\qty[\eta(\omega)] < 0$. However, there is an intrinsic difference between a change in space and a change in time: one can only allow for an evolution in the positive $t$ direction, while evolution both in the positive and in the negative $x$ direction can be allowed. As already pointed out in Section~\ref{spaevo}, this physical difference between the variables $t$ and $x$ is due to the causality principle which holds in time, but not in space. Therefore, the assessment of growth or damping along the $x$ axis is not important, but one must check if the signal amplitude either grows or is damped along its direction of propagation. One can set the angular frequency of the wave, $\omega$, to be positive or zero, without any loss of generality. Thus, the direction of propagation is established either by the sign of the phase speed, $\omega/k$, or by the sign of the wavenumber, $k$. 

Let us write
\eqn{
\eta = s + i\, k ,
}{29}
where $s = \Re\!\qty(\eta)$ is the spatial growth rate and $k=\Im\!\qty(\eta)$ is the wavenumber.
We define spatial stability a condition where
\eqn{
s\,k < 0 .
}{29ant1}
On the other hand, spatial instability is defined by the condition
\eqn{
s\,k > 0.
}{29ant2}
One may point out that, excluded from \equasa{29ant1}{29ant2}, there may exist a special case where $k=0$. Such a case depicts a situation where there are no oscillations in space, but we may have oscillations in time since the angular frequency is assumed to be positive or zero. In this special case, $s \ne 0$ always entails spatial instability either in the positive $x$ direction (if $s>0$) or in the negative $x$ direction (if $s<0$). 

\begin{figure}
\centering
\includegraphics[width=0.4\textwidth]{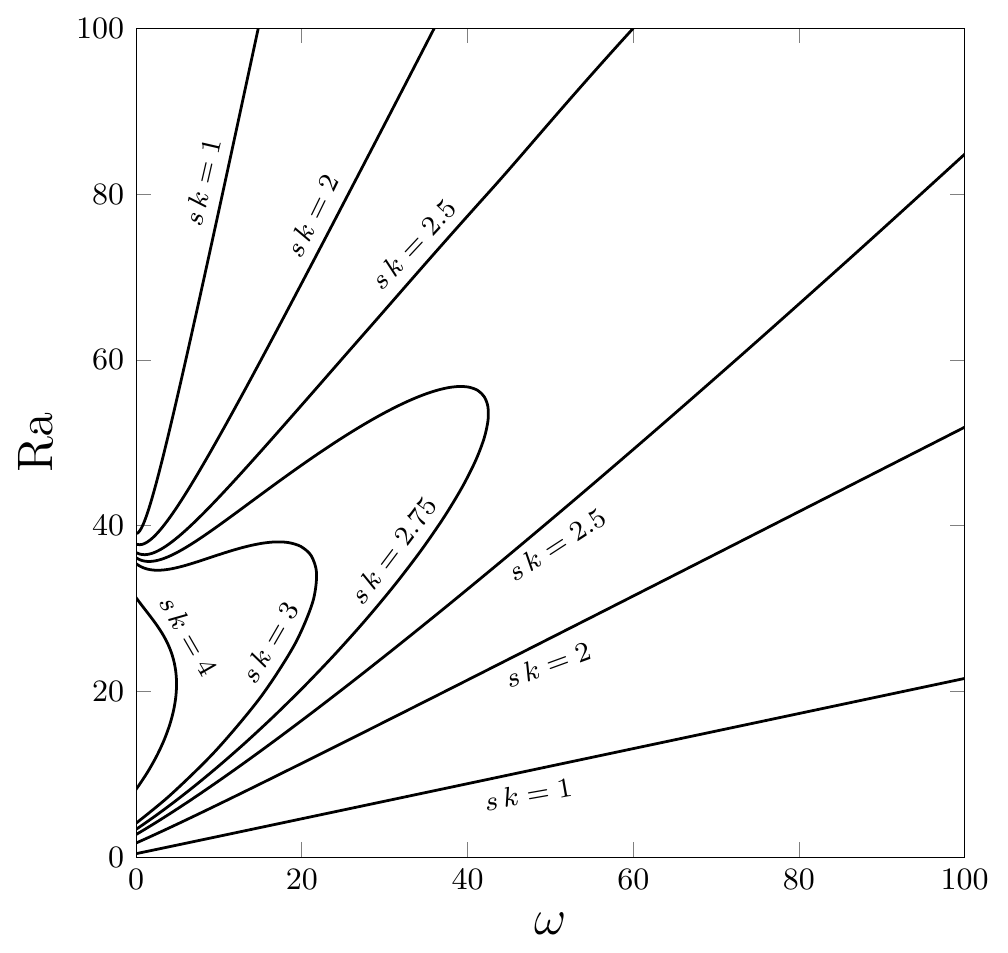}
\caption{\label{fig2}Plots of the isolines of $s\,k$ in the $(\omega,\ra)$ plane for $\pec=0$ and conditions of spatial instability $\qty(s\,k > 0)$.}
\end{figure}

The marginal condition where $s = 0$ is also excluded from \equasa{29ant1}{29ant2}. This case defines the intersection between the set of perturbation modes allowed by the definition of Fourier transform implied by \equasa{14}{15} and the set of Fourier modes defined by \equasa{23}{24}. In fact, the marginal condition given by $s=0$ identifies Fourier modes which may oscillate in space and in time, but whose amplitude is constant both in space and in time. In the context of Section~\ref{timdev}, such modes are those defining the neutral stability condition while, in the present context of time--periodic modes, they definitely yield a condition of spatial stability, albeit marginal, as there is no damping in space.
If $s=0$, \equa{28} yields
\eqn{
\ra = \frac{\qty[k^2 + \qty(n \pi)^2]^2}{k^2} \qand \omega = k\, \pec ,
}{30}
which means
\eqn{
\ra = \frac{\qty[\omega^2 + \qty(n \pi)^2 \pec^2]^2}{\omega^2\, \pec^2} .
}{31}
\Equasa{30}{31} convey an important result. The marginal condition $s=0$ is achieved at its lowest by setting $n=1$. Then, it has exactly the same form as the threshold to convective instability within the analysis of time--evolving modes presented in Section~\ref{timdev}. This is easily established on comparing \equa{30} with \equasa{18}{20}. 

The implicit form of the dispersion relation (\ref{28}) cannot be encompassed in the case where $s \ne 0$. There are, in fact, four branches $\eta(\omega)$ which can be exploited by solving \equa{28}, for given $n$ and $\pec$. 

An important feature of \equa{28} is its dependence on $n$. We can employ the scaling
\eqn{
\eta'=\frac{\eta}{n} \qc \omega'=\frac{\omega}{n^2} \qc \ra' = \frac{\ra}{n^2} \qc \pec' = \frac{\pec}{n}.
}{32}
With such a scaling, \equa{28} can be rewritten as
\eqn{
\eta'\,^2 - \pi^2 + i \omega' + \frac{\eta'\,^2\, \ra'}{\eta'\,^2 - \pi^2} - \eta'\, \pec' = 0 ,
}{33}
which formally coincides with \equa{28} for the special case $n=1$. Thus, we can encompass the dependence on $n$ by employing the primed quantities. Hereafter, the primes will be omitted, for the sake of simplicity in the notation, but we will keep in mind that any special value of $\qty(\eta, \omega, \ra, \pec)$ must be implicitly scaled proportionally to $\qty(n, n^2, n^2, n)$.

\subsection{Darcy--B\'enard system $(\pec=0)$}\label{darbensys}
\Equa{28} has four roots $\eta = \eta_j$, with $j=1, \ldots, 4$. Although, in general, the analytical expression of such roots is quite complicated, they can be easily written in the case when either the basic flow rate vanishes or the basic flow direction is parallel to the $x$ axis, {\em i.e.}, when $\pec = 0$. In fact, one can write 
\spl{
\eta_1 = \sqrt{\frac{1}{2}\, \qty[2 \pi^2 - \ra - i \omega - \sqrt{\qty(\ra + i \omega)^2 - 4 \pi^2 \ra}\,]} ,
\nonumber\\
\eta_2 = -\,\eta_1,
\nonumber\\
\eta_3 = \sqrt{\frac{1}{2}\, \qty[2 \pi^2 - \ra - i \omega + \sqrt{\qty(\ra + i \omega)^2 - 4 \pi^2 \ra}\,]} ,
\nonumber\\
\eta_4 = -\,\eta_3.
}{34}
Figure~\ref{fig1} illustrates the behaviour of the real part $s$ of the four roots $\eta_j$ given by \equa{34} as $\omega$ increases, for prescribed values of $\ra$. The symmetry with respect to the $\omega$ axis of the plots reported in Fig.~\ref{fig1} is evident and it is a consequence of \equa{28} being dependent on $\eta^2$ when $\pec=0$.

\begin{figure*}
\centering
\includegraphics[width=0.8\textwidth]{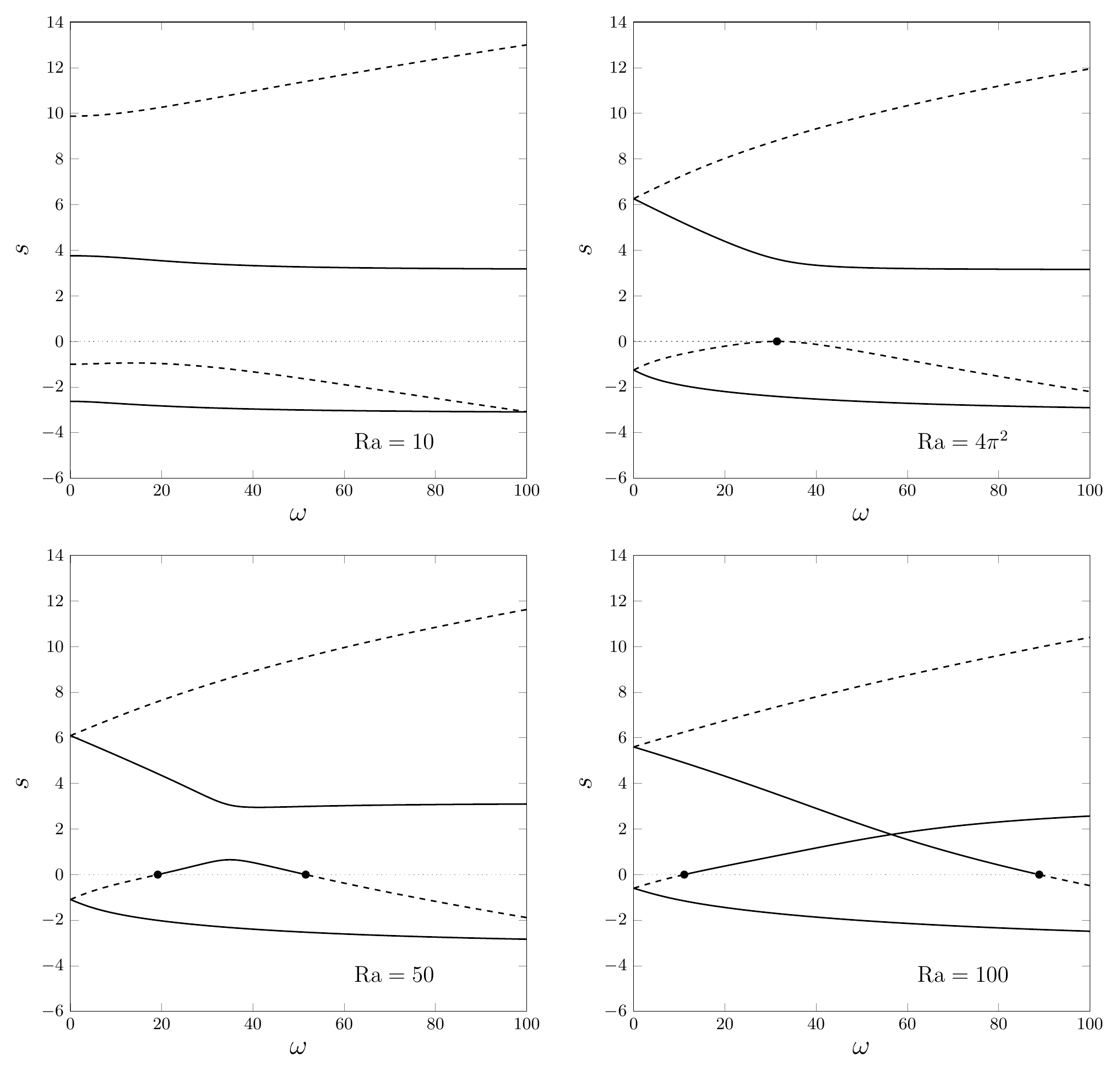}
\caption{\label{fig3}Plots of $s$ versus $\omega$ for $\pec=10$ and different values of $\ra$. Dashed lines denote spatial stability $\qty(s\,k < 0)$, while solid lines denote spatial instability $\qty(s\,k > 0)$.}
\end{figure*}

Figure~\ref{fig1} illustrates the behaviour of the spatial growth rate $s$ for the subcritical case $\ra=10$, for the critical case $\ra=4\pi^2$, and for the two supercritical cases $\ra=50$ and $100$. Spatially stable $\qty(s\,k < 0)$ branches are drawn as dashed lines, while spatially unstable $\qty(s\,k > 0)$ branches are drawn as solid lines. The unstable branches are those associated with the roots $\eta_3$ and $\eta_4$ defined by \equa{34}.
The remarkable fact about Fig.~\ref{fig1} is that, whatever is the positive value of $\ra$ either subcritical, critical or supercritical, there are always spatially unstable modes. This means that the Darcy--B\'enard system is always spatially unstable. Furthermore, the special condition described by \equa{31} and defining the locus $s=0$ yields an infinite $\ra$ when $\pec \to 0$ with $\omega\ne 0$. 

\begin{figure*}
\centering
\includegraphics[width=0.8\textwidth]{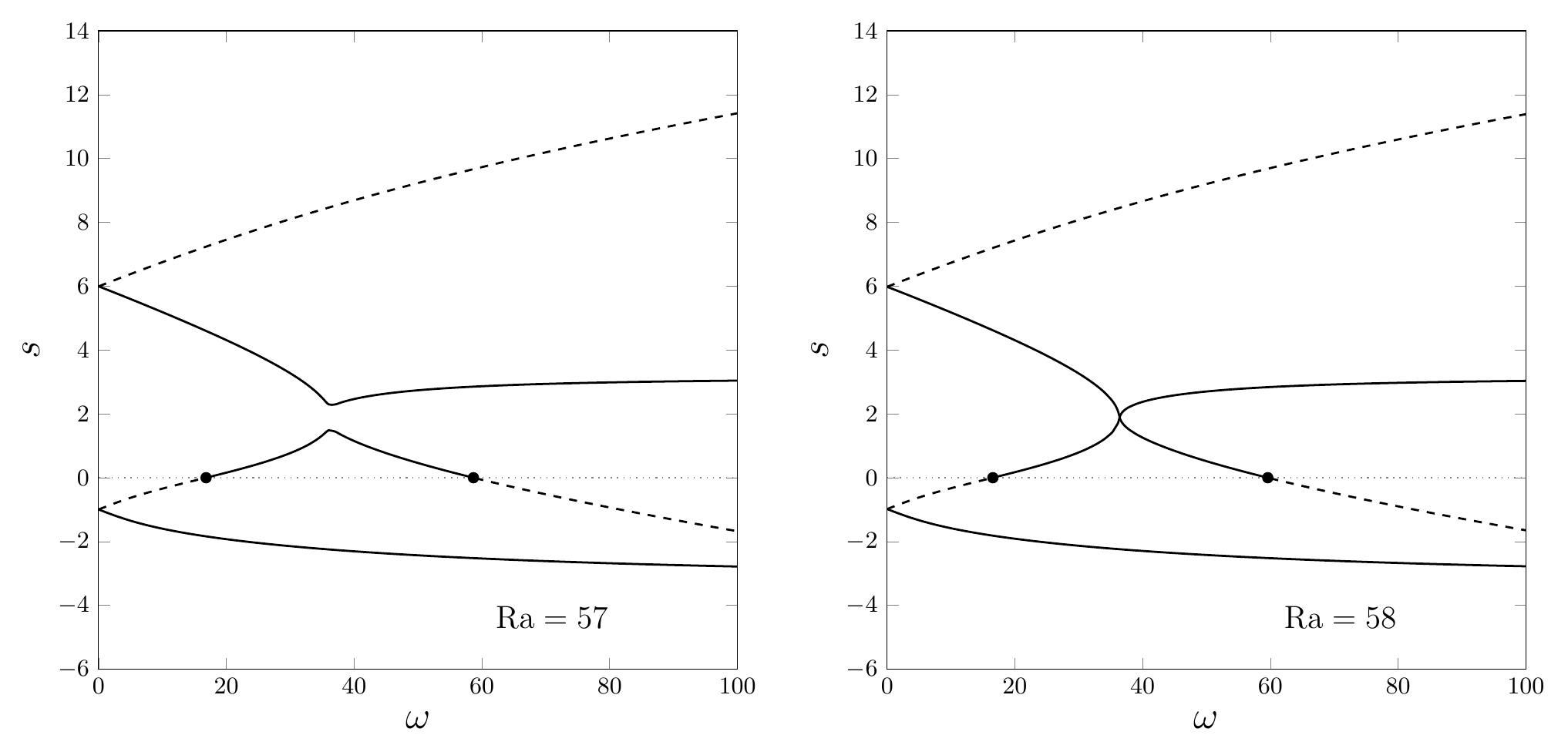}
\caption{\label{fig4}Plots of $s$ versus $\omega$ for $\pec=10$ and $\ra=57$ and $58$. Dashed lines denote spatial stability $\qty(s\,k < 0)$, while solid lines denote spatial instability $\qty(s\,k > 0)$.}
\end{figure*}

\begin{figure*}
\centering
\includegraphics[width=0.8\textwidth]{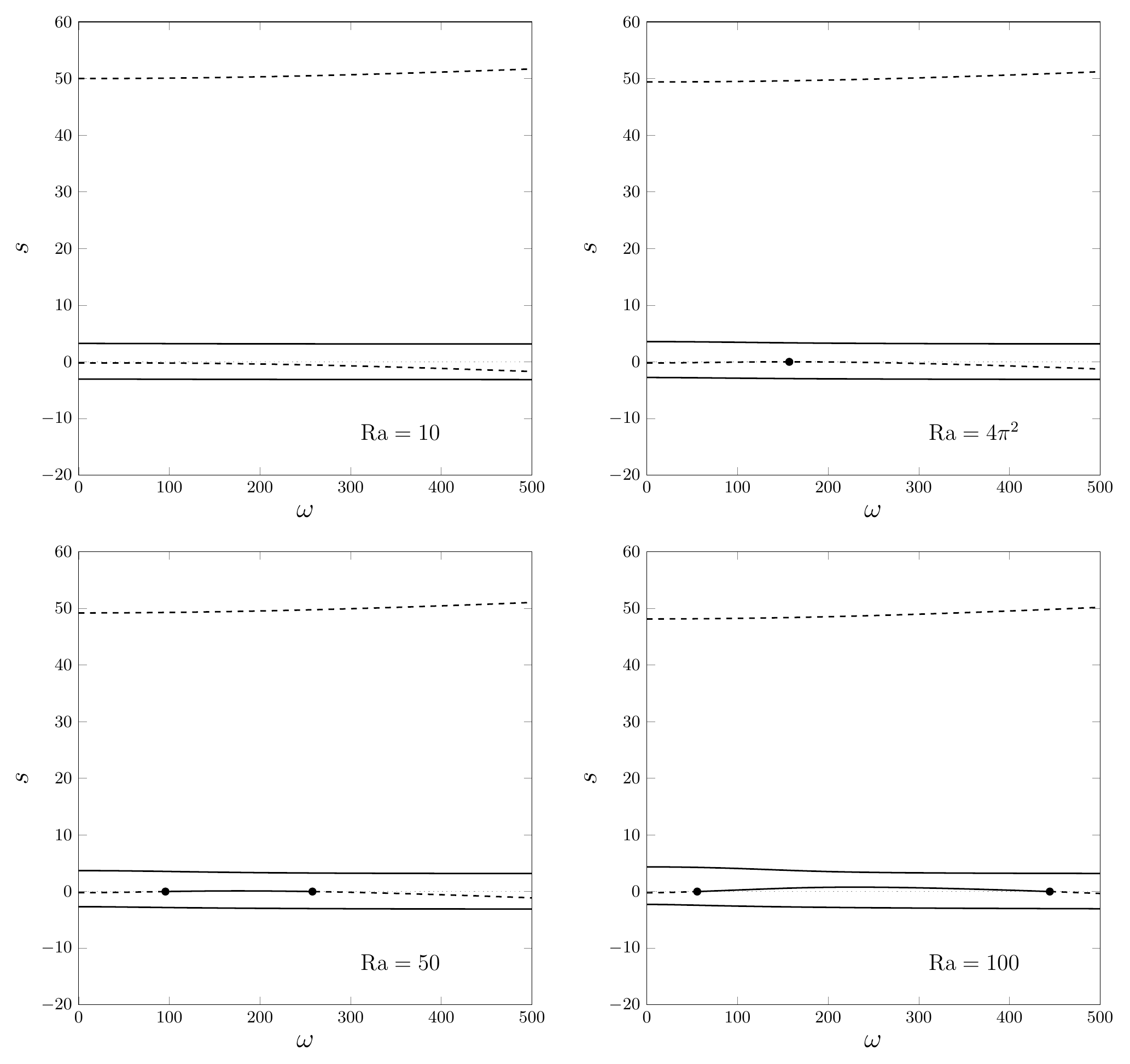}
\caption{\label{fig5}Plots of $s$ versus $\omega$ for $\pec=50$ and different values of $\ra$. Dashed lines denote spatial stability $\qty(s\,k < 0)$, while solid lines denote spatial instability $\qty(s\,k > 0)$.}
\end{figure*}

The whole parametric plane $(\omega, \ra)$ includes modes with $\qty(s\, k > 0)$. This circumstance is illustrated in Fig.~\ref{fig2} where the lines with $s\,k =constant$ are drawn in the $(\omega, \ra)$ plane for conditions of spatial instability. Such isolines can be easily drawn as parametric plots by employing the expressions {of $\ra$ and $\omega$} as functions of $k$ which can be gathered from \equa{28} with $\pec=0$,
{
\spl{
\ra = \frac{\left[k^2 + (s - \pi)^2\right] \left(k^2-s^2+\pi ^2\right) \left[k^2+(s+\pi )^2\right]}{k^4+k^2 \left(2 s^2+\pi ^2\right)+ s^2 \qty(s^2 - \pi^2)} ,
\nonumber\\
\omega = -\frac{2 k s \left[\left(k^2+s^2\right)^2-\pi ^4\right]}{k^4+k^2 \left(2 s^2+\pi ^2\right)+ s^2 \qty(s^2 - \pi^2)} .
}{35}
}
A comprehensive discussion of the spatial instability in the Darcy--B\'enard system can be found in \citet{barletta2021spatially}.

\begin{figure}
\centering
\includegraphics[width=0.4\textwidth]{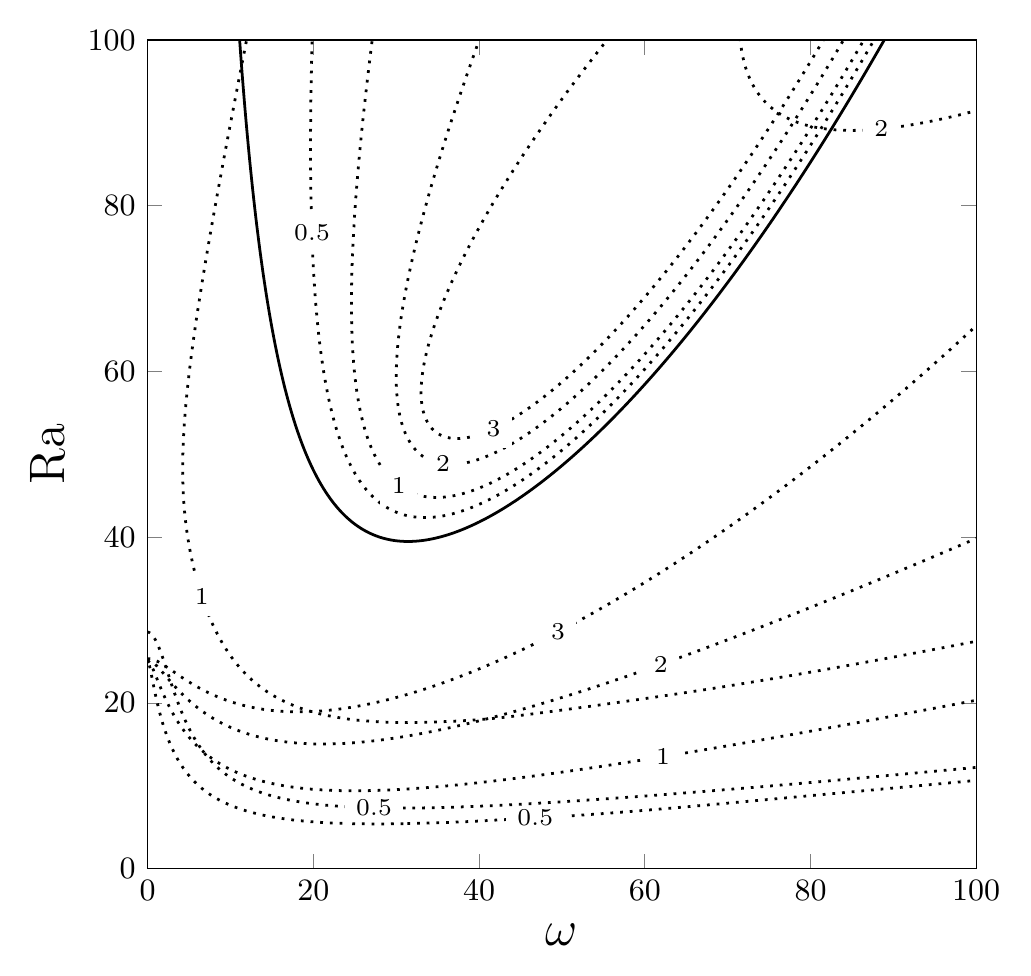}
\caption{\label{fig6}Plots of the isolines of $s\,k$ (dotted lines) with $s\,k > 0$ in the $(\omega,\ra)$ plane for $\pec=10$. The solid line denotes the neutral stability curve $(s=0)$.}
\end{figure}

\begin{figure}
\centering
\includegraphics[width=0.4\textwidth]{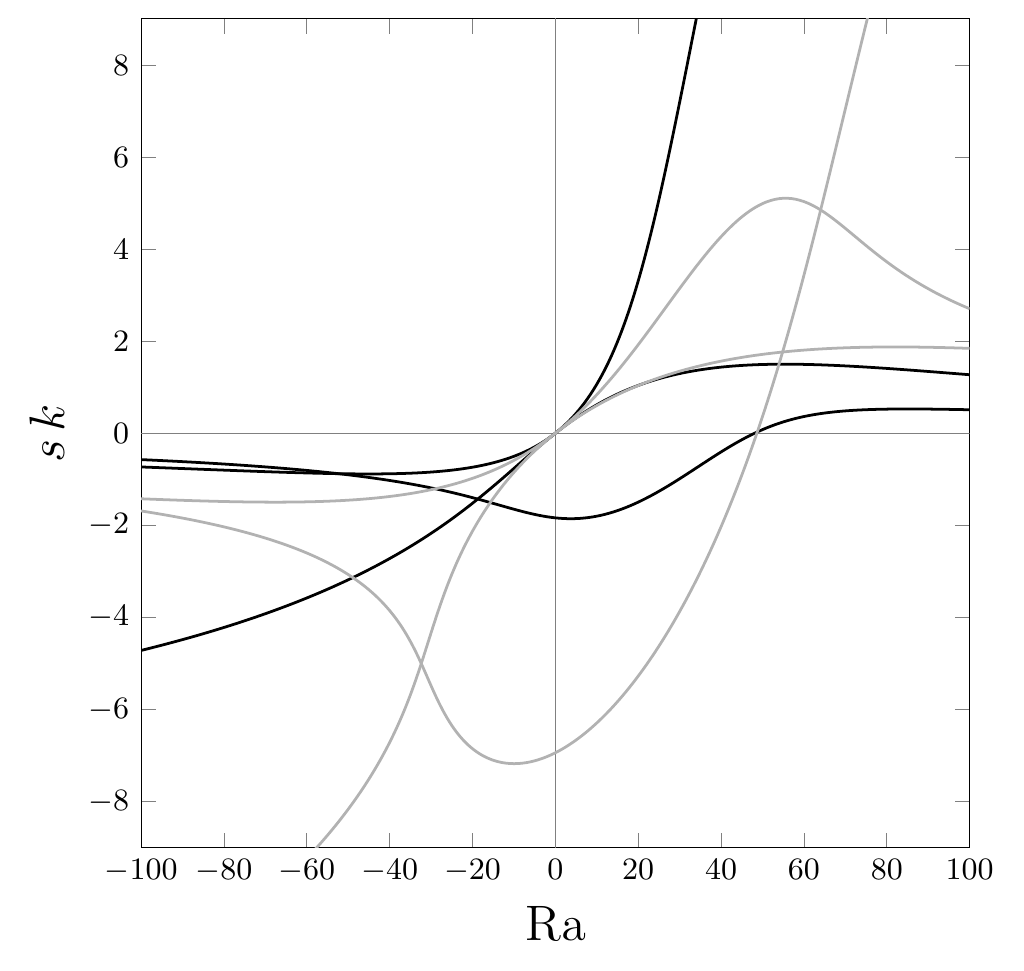}
\caption{\label{fig7}Plots of of $s\,k$ versus $\ra$ for $\pec=10$ and either $\omega=20$ (black lines) or $\omega=50$ (grey lines).}
\end{figure}

\subsection{Non--vanishing flow rate $(\pec > 0)$}
The symmetry with respect to the $\omega$ axis discussed with reference to Fig.~\ref{fig1} is lost when $\pec > 0$. This feature emerges quite clearly in Figs.~\ref{fig3}--\ref{fig5}. In these figures, plots of $s$ versus $\omega$ are provided for different values of $\ra$ and $\pec=10$ or $\pec=50$. Again, the dashed lines entail spatial stability conditions $\qty(s\,k < 0)$, while solid lines connote spatial instability $\qty(s\,k > 0)$.

For every given pair $(\pec, \ra)$ one has both positive and negative branches of $s$ which, in some cases. cross the $\omega$ axis and change their sign. Figures~\ref{fig3} and \ref{fig4} are relative to $\pec=10$, while Fig.~\ref{fig5} is for the case $\pec=50$. Black dots are drawn in these figures to identify the points where $s=0$, namely the intersections with the $\omega$ axis. Such points can be easily detected by solving \equa{31} with given $\ra$ and $\pec$. It is quite expected that the condition $s=0$ emerges only when $Ra \ge Ra_c = 4\pi^2$. In fact, the minimum $\ra$ is attained, according to \equa{31}, if $\omega=\omega_c = \pi\, \pec$. 
Another interesting feature displayed in Fig.~\ref{fig3} is that, for the supercritical values $\ra=50$ and $\ra=100$, the two central branches $s\qty(\omega)$ in the plot gradually approach and, eventually, they meet (frame with $\ra=100$). Figure~\ref{fig4} shrinks the interval of $\ra$ where the two central branches $s(\omega)$ approach and clap together. In Fig.~\ref{fig4}, it is evident that the central branches merge for a value of $\ra$ within $57$ and $58$ and for a value of $\omega$ within $30$ and $40$. The exact values are, in fact,
\eqn{
\ra = 57.8036 \qc \omega = 36.2947 \qc s = 1.89300 .
}{36}
Interestingly enough, such values coincide with those characterising the transition to absolute instability when $\pec=10$, as reported in \citet{barletta2019routes}. This is not accidental given that the branches merging is accompanied, as evidently shown in the frame of Fig.~\ref{fig4} for $\ra = 58$, by a situation where the derivative $\dd s\qty(\omega)/\dd \omega$ becomes infinite, which is equivalent to having $\dd \omega/\dd s = 0$. The latter condition is a restatement, for time--periodic Fourier modes, of the definition of saddle point, $\dd \lambda/\dd k = 0$, which entails the transition to absolute instability as mentioned in Section~\ref{absinstity} and extensively discussed in \citet{barletta2019routes}. 

Incidentally, we point out that the merging of the central branches in the $(\omega, s)$ plane is present also for $\pec=0$, as illustrated in Fig.~\ref{fig1}. In this case, the merging occurs at the origin, $\omega=0$, when $\ra = \ra_c = 4\pi^2$ and with $s=0$. This result is perfectly coherent with the transition to absolute instability taking place, with $\pec=0$, at the onset of convective instability \citep{barletta2019routes}. No merging between the central branches $s\qty(\omega)$ is displayed in Fig.~\ref{fig5}. This is not surprising since, for $\pec=50$, the transition to absolute instability happens at $\ra = 208.441$, as reported by \citet{barletta2019routes}, which is outside the range considered in Fig.~\ref{fig5}. 

\begin{figure}
\centering
\includegraphics[width=0.4\textwidth]{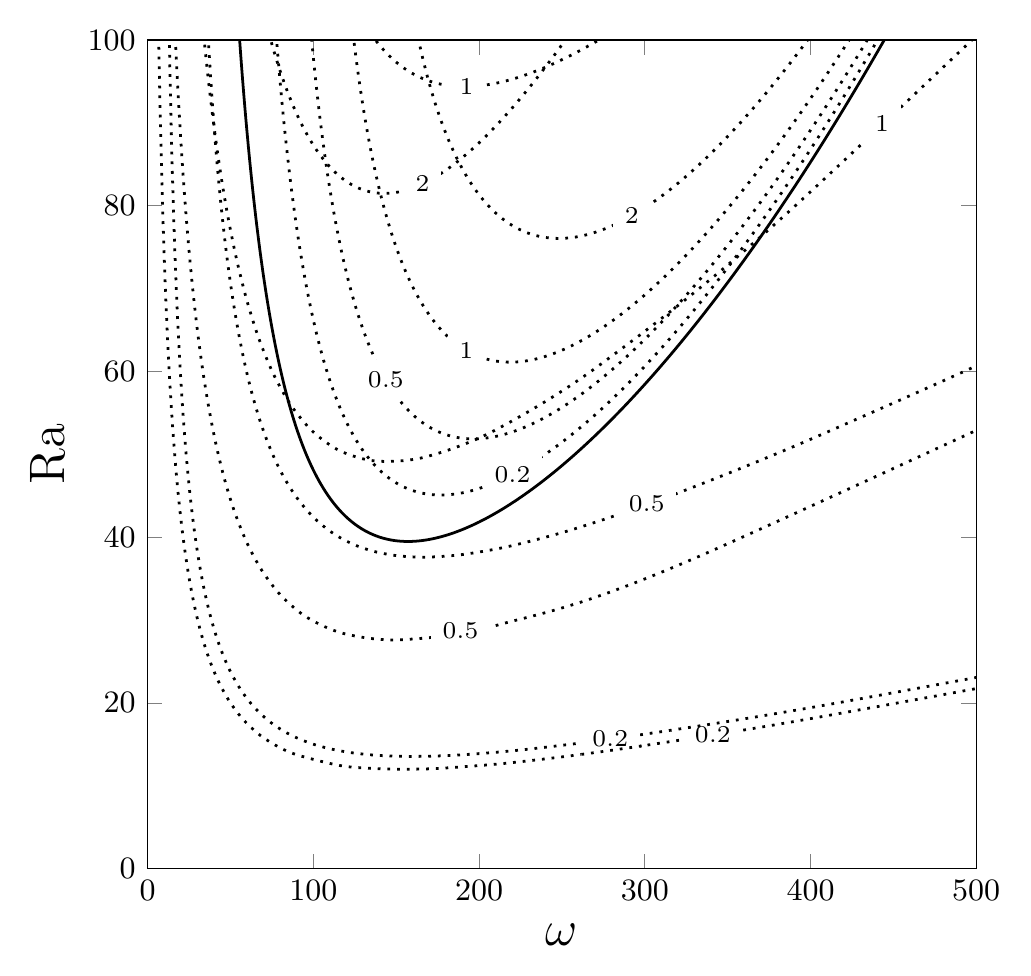}
\caption{\label{fig8}Plots of the isolines of $s\,k$ (dotted lines) with $s\,k > 0$ in the $(\omega,\ra)$ plane for $\pec=50$. The solid line denotes the neutral stability curve $(s=0)$.}
\end{figure}

\begin{figure}
\centering
\includegraphics[width=0.4\textwidth]{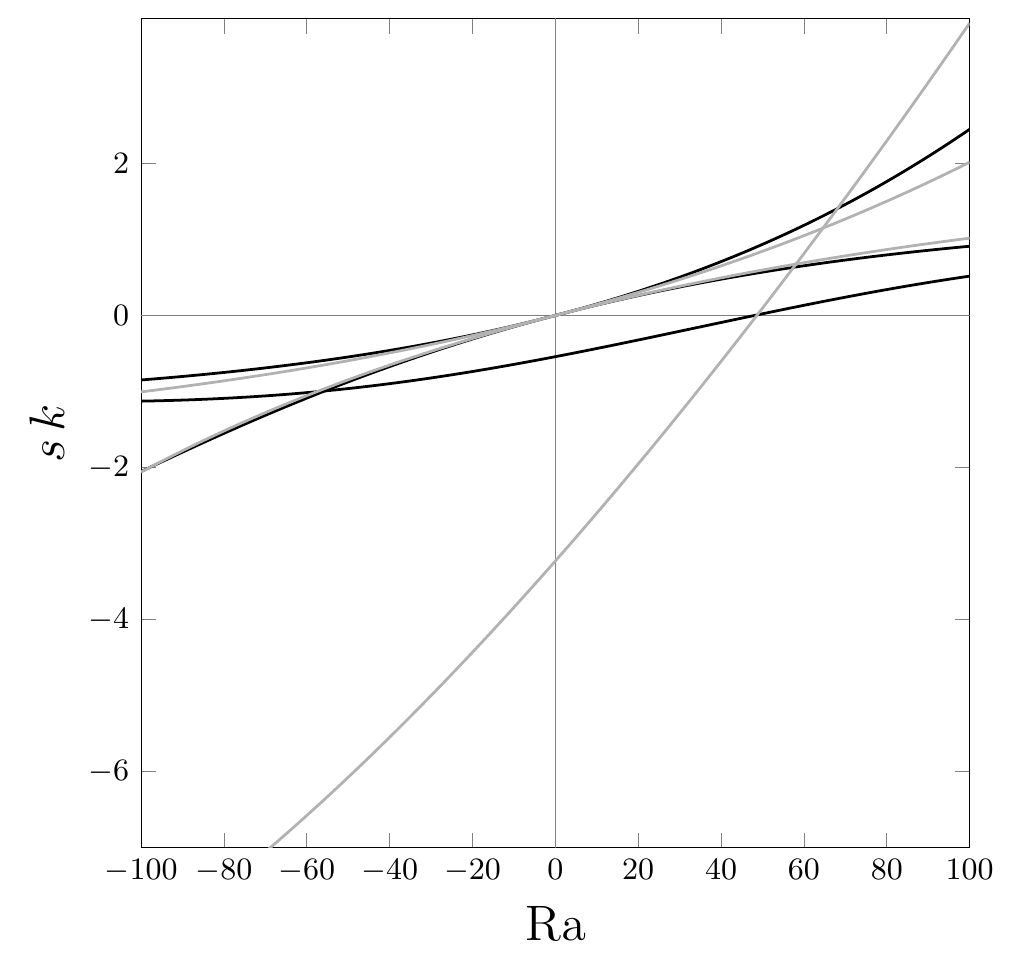}
\caption{\label{fig9}Plots of of $s\,k$ versus $\ra$ for $\pec=50$ and either $\omega=100$ (black lines) or $\omega=250$ (grey lines).}
\end{figure}

From Figs.~\ref{fig3}--\ref{fig5}, it is clear that there are always positive and negative values of the spatial growth rate $s$, exactly as for the Darcy--B\'enard problem examined in Section~\ref{darbensys}. 
The physical rationale is exactly the same. There always exist spatially developing modes which can destabilise the basic solution either in the rightward $x$ direction $(x>0)$ or in the leftward $x$ direction $(x<0)$. As we are envisaging cases with $\pec \ge 0$, we conventionally call upstream region the domain $x>0$ and upstream region the domain $x<0$.
What happens in the upstream region can be described by recognising that the most unstable modes are those with the smallest negative $s$. The amplification upstream is, again, always present although its rate of exponential change in the negative $x$ direction is, with a few exceptions (see, for instance, Fig.~\ref{fig4} for the case $\ra = 58$), smaller than that occurring downstream. 
A counterexample for the case $\ra=58$, illustrated in Fig.~\ref{fig4}, is for $\omega = 36$. With this angular frequency, for the spatially unstable modes, the maximum spatial growth rate downstream is $s=2.159$, while the maximum growth rate upstream is $\qty|s|=2.234$. 
This feature is yet another way to interpret the lack of symmetry with respect to the $\omega$ axis in all the plots reported in Figs.~\ref{fig3}--\ref{fig5}. In fact, the symmetry observed in Fig.~\ref{fig1} for $\pec = 0$ implies that the amplification of the perturbation modes upstream $(x<0)$ always coincide with that downstream $(x>0)$. With $\pec \ne 0$, this symmetric behaviour disappears. 

We have considered $\pec \ge 0$ in all examples so far, but taking $\pec \le 0$ does not change the physical interpretation in any substantial way. Indeed, we have just to consider the negative $x$ direction as downstream and the positive $x$ direction as upstream.

Figures~\ref{fig6} and \ref{fig7} are relative to $\pec=10$. They illustrate how spatially unstable modes $(s\,k > 0)$ occur for every $\ra > 0$, while they do not emerge in the case of stable thermal stratification, $\ra < 0$. The latter feature is quite important as it shows that stable thermal stratification means both Lyapunov's linear stability and spatial stability. Figure~\ref{fig6} depicts the $(\omega, \ra)$ plane with the neutral stability curve given by \equa{31} (solid line). The dotted lines are the isolines of $s\,k$ with positive values $0.5, 1, 2$ and $3$. Such spatially unstable isolines are spread across the $(\omega, \ra)$ plane both above and below the neutral stability curve. This, once again, means that spatial instability arises both in subcritical and in supercritical conditions. On the other hand, spatial instability is ruled out for $\ra < 0$ as shown by Fig.~\ref{fig7}. In fact, this figure shows the possible branches of $s\,k$ versus $\ra$ with either $\omega=20$ (black lines) or $\omega=50$ (grey lines). In these cases, $s\, k$ cannot be positive with a negative $\ra$, which is a general result. This means that spatial instability is out of the question when $\ra < 0$.

The data employed in Fig.~\ref{fig6} are evaluated by employing the generalised version of \equa{35}, namely
%
\spl{
\ra = \frac{\qty[k^2 + (s - \pi)^2] \qty[k^2 + ( s + \pi )^2] \qty[k^2 + s (\pec - s) + \pi^2 ]}{k^4 + k^2 \qty(2 s^2 + \pi^2) + s^2 \qty(s^2 - \pi^2)} ,
\nonumber\\
\omega = \frac{k \left(k^2 + s^2 + \pi ^2\right) \qty[\qty( k^2  + s^2 ) \qty(\pec - 2 s) + 2 s \pi^2)]}{k^4 + k^2 \qty(2 s^2 + \pi^2) + s^2 \qty(s^2 - \pi^2)} .
}{37}
%
In fact, we recall that \equa{35} holds just for the case $\pec=0$, while \equa{37} is valid for every $\pec$.

The test carried for $\pec=10$ in Figs.~\ref{fig6} and \ref{fig7} is repeated for $\pec=50$ in Figs.~\ref{fig8} and \ref{fig9}. The behaviour is just the same observed for a smaller $\pec$: there is spatial instability both above and below the neutral stability curve. Furthermore, by exploring the cases $\omega=100$ and $\omega=250$ in Fig.~\ref{fig9}, we do not detect any spatial instability when $\ra<0$. Indeed, there always exist spatial modes growing in either the downstream or the upstream direction, whatever is the value of $\omega$ and the (positive) value of $\ra$. Roughly speaking, this is equivalent to saying that the linear spatial instability of the basic flow (\ref{5}) always occurs, no matter how small is the Rayleigh number, provided that $\ra > 0$.

\section{Conclusions}
The linear instability of the stationary throughflow in a horizontal porous channel, known in the literature as Prats' problem, has been analysed by employing time--periodic modes whose amplitude may vary along a given horizontal direction. This approach, called spatial stability analysis, sheds a new light on the conditions leading to a linearly unstable behaviour for Prats' problem. In fact, the widely accepted result is that a linearly unstable behaviour of the horizontal throughflow arises when the Rayleigh number, $\ra$, becomes larger than its critical value, $4\,\pi^2$. This result, which holds for every value of the modified P\'eclet number, $\pec$, relies on the classical stability analysis based on space--periodic and time--evolving modes. 
The classical analysis of linear stability is, in fact, an application of Lyapunov's definition of instability, where the time--evolution of the dynamical system is tested versus small perturbations of its initial state. On the other hand, the spatial stability approach monitors the effects in the flow domain of a time--periodic source acting at a given spatial position. The applied periodic forcing can be either amplified or damped along a given spatial direction. The spatially unstable behaviour occurs when a given Fourier mode of perturbation is amplified along its direction of propagation. In this paper, these concepts have been introduced step--by--step, thus leading to the following results:
\\[-16pt]
\begin{itemize}
\item For every given pair $\qty(\pec, \ra)$, there are four branches of Fourier perturbation modes which are time--periodic, with an angular frequency $\omega$, and which are endowed with a complex growth--rate, $\eta$. The real part of $\eta$, denoted with $s$, is the spatial growth rate along a horizontal axis, $x$. The imaginary part of $\eta$, called $k$, is the wavenumber of the mode.
\\[-16pt]
\item A regime of spatial instability is defined when the product $s\, k$ is positive.
\\[-16pt]
\item The basic throughflow along the porous channel is spatially unstable for every positive value of the Rayleigh number. A regime where $\ra > 0$ defines a condition of heating from below.
\\[-16pt]
\item The spatial instability has been explicitly investigated by considering the values $\pec = 10$ and $\pec=50$. Significant geometrical features of the spatial modes are detected when the Rayleigh number equals its critical value $4\,\pi^2$, as well as its threshold value for the transition to the absolute instability. In particular, the transition to absolute instability is accompanied by the merging of two spatial instability branches in the plane where $s$ is drawn versus $\omega$. The merging of these branches happens with an infinite derivative, $d s/d \omega$, which can be identified with the saddle--point condition typical of the absolute instability threshold.
\end{itemize}
The results obtained in this paper offer a new perspective for the concept of unstable flow in a porous channel. There are opportunities for an experimental validation, by testing the streamwise spatial evolution, both upstream and downstream, induced by a localised time--periodic source of perturbations. There are also significant interesting developments that can be obtained by devising different flow regimes and boundary conditions for the porous channel. Future research can also address the role played by the nonlinearity of convection heat transfer. This task can be accomplished by investigating the actual nonlinear, or weakly nonlinear, flow regions downstream and upstream of a localised harmonic source of temperature/velocity perturbations. In fact, one reasonably expects that the exponential spatial growth predicted by the linear analysis is tamed by a nonlinear saturation effect, just as it happens for the classical analysis of instability to temporal modes.

\section*{Acknowledgements}
The author acknowledges the financial support from the grant PRIN 2017F7KZWS provided by the Italian Ministry of Education and Scientific Research.

\section*{Data availability}
The data that supports the findings of this study are available within the article.


\section*{References}

\providecommand{\BIBde}{de~B}

\appendix
\section{A toy model of spatial modes}\label{App}
Let us consider the one--dimensional heat conduction in a semi--infinite solid medium occupying the region $x>0$. The heat diffusion process is described through Fourier's equation,
\eqn{
\pdv{T}{t} = \alpha \, \pdv[2]{T}{x},
}{AA1}
where $\alpha$ is the thermal diffusivity of the solid. Let us assume that a steady--periodic temperature signal is supplied at $x=0$,
\eqn{
T(0,t) = A\, \cos\qty(\omega \, t),
}{AA2} 
where $A$ is a constant and $\omega$ is the angular frequency. 

\begin{figure}[h]
\centering
\includegraphics[width=0.4\textwidth]{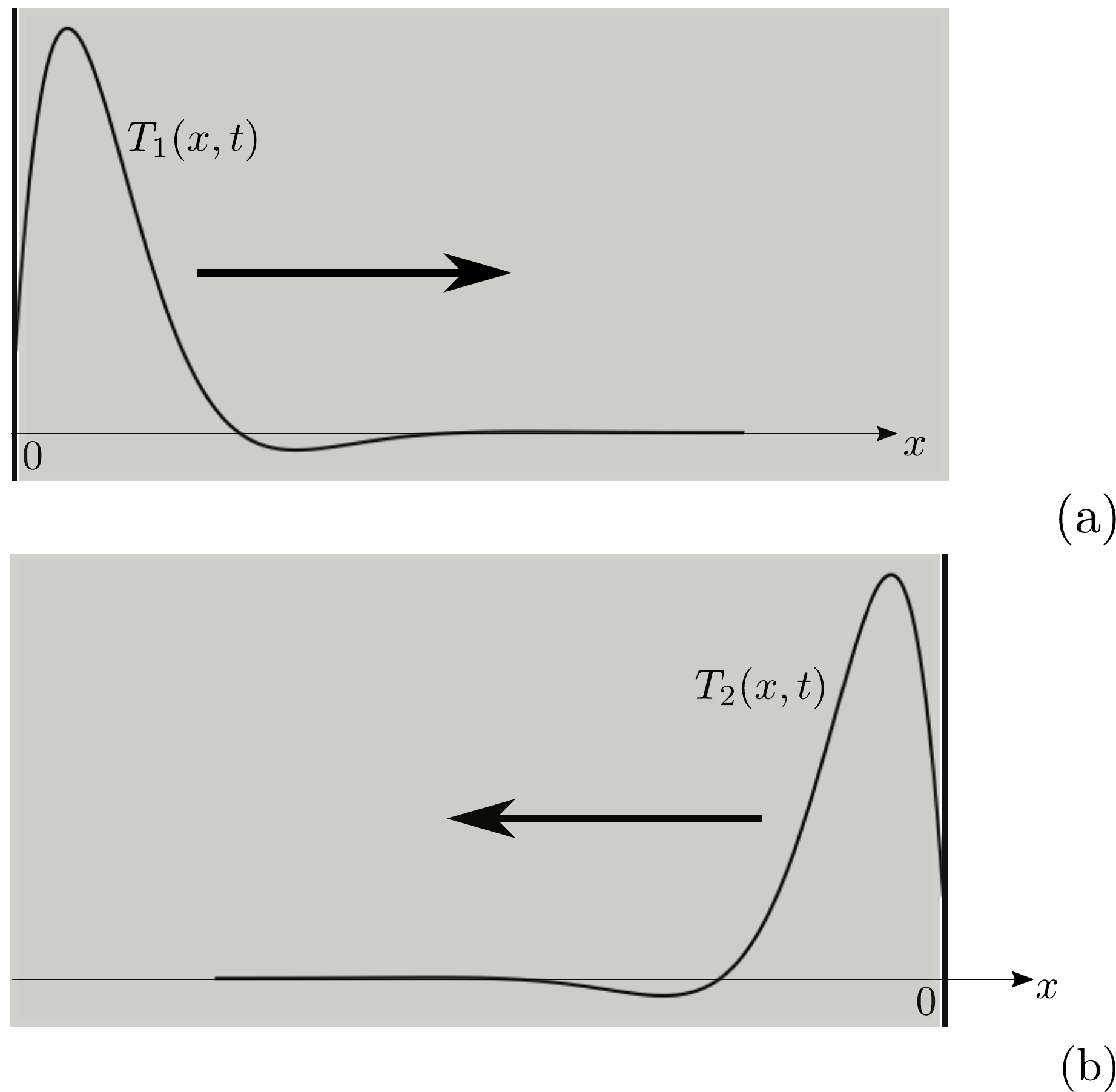}
\caption{\label{fig10}Qualitative sketch of the solutions $T_1(x,t)$ and $T_2(x,t)$, at a fixed time $t$, for a semi--infinite solid medium extending over $x>0$ (a), and for its dual filling the region $x<0$ (b).}
\end{figure}

\begin{figure}[h]
\centering
\includegraphics[width=0.4\textwidth]{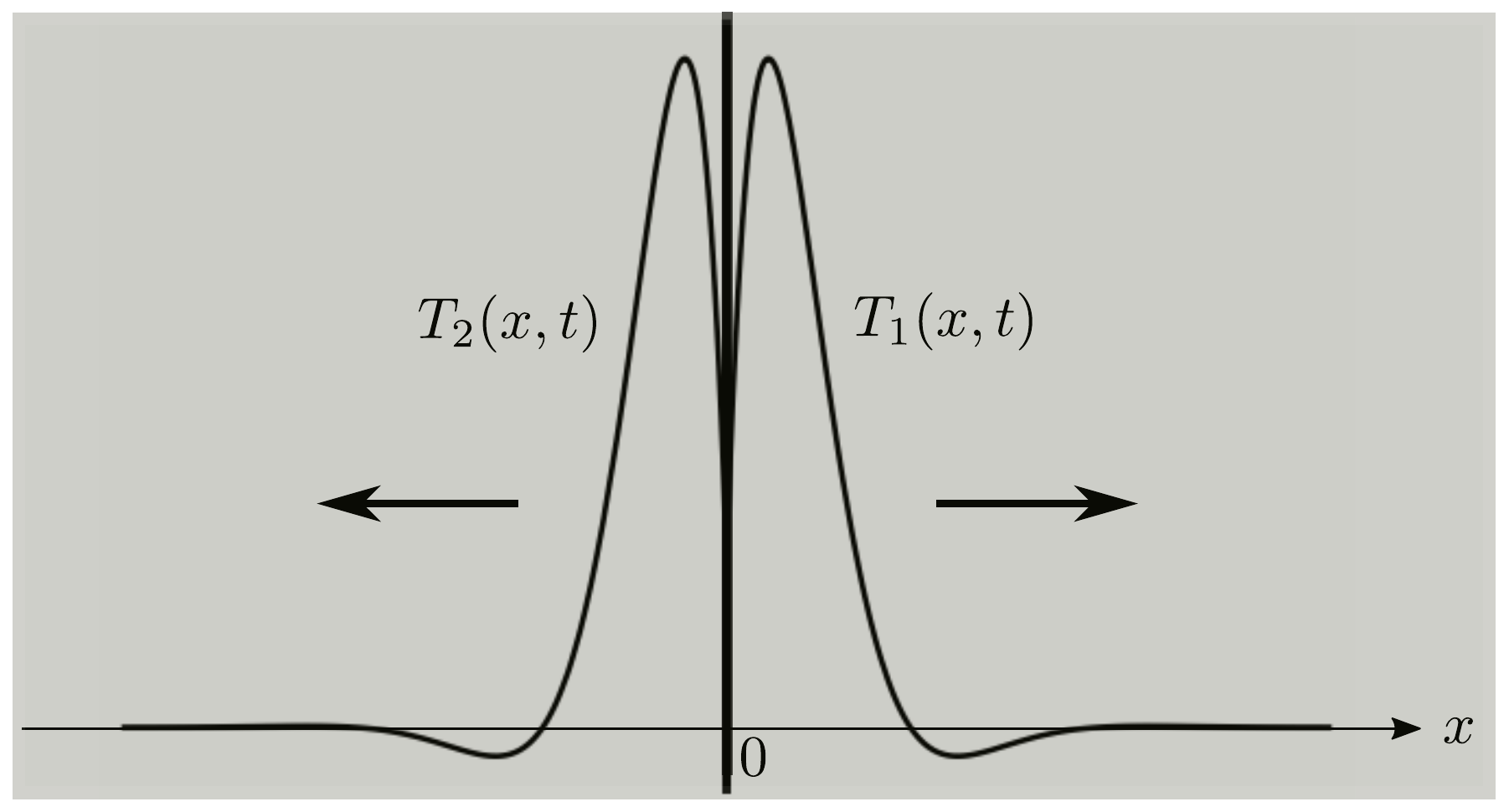}
\caption{\label{fig11}Qualitative sketch of the solutions $T_1(x,t)$ and $T_2(x,t)$, at a fixed time $t$, for an infinite solid medium extending over $-\infty < x < +\infty$, with a localised harmonic source at $x=0$.}
\end{figure}

In the steady--periodic regime, there exist two possible solutions for \equasa{AA1}{AA2},
\gat{
T_1(x,t) = A \exp\!\qty({-x \sqrt{\frac{\omega}{2 \alpha}}}\;) \cos\!\qty(\omega\, t - x  \sqrt{\frac{\omega}{2 \alpha }}\;) ,\label{AA3a}\\
T_2(x,t) = A \exp\!\qty({x \sqrt{\frac{\omega}{2 \alpha}}}\;) \cos\!\qty(\omega\, t + x  \sqrt{\frac{\omega}{2 \alpha }}\;) .\label{AA3b}
}{AA3}
This is what mathematics says about the temperature distribution in the semi--infinite solid with harmonic temperature forcing at $x=0$. 

The question is: which is the correct exploitation of the actual temperature response in the semi--infinite medium: \equa{AA3a} or \equa{AA3b}? It is worth being mentioned that \equasa{AA1}{AA2} represent an oversimplified model of a real--world situation, namely the temperature change in time, at a given depth, inside the soil. Here, $x >0$ is precisely the depth where the temperature is evaluated. 

The harmonic forcing at the ground surface $x=0$ could be caused, for instance, by the seasonal change of the air temperature. 
It is utterly evident that the correct physical solution is given by \equa{AA3a}. Indeed, $T_1(x,t)$ represents a wave propagating in the positive $x$ direction and damped for increasing values of $x$. On the other hand, $T_2(x,t)$ represents a wave propagating in the negative $x$ direction and amplified for increasing values of $x$. 
If the solid medium occupies the semi--infinite region $x>0$, there is no chance to observe a temperature signal amplification for increasing values of the depth $x$ such as that predicted by $T_2(x,t)$. Thus, one could question on the reasons why mathematics leads to a ghost solution, $T_2(x,t)$, for a physical problem. Indeed, the model defined by \equasa{AA1}{AA2} does not contain any information about the domain where the solution is to be employed. 

As illustrated in Fig.~\ref{fig10}, \equa{AA3b} provides the solution for the dual problem where the semi--infinite solid medium extends over the region $x<0$. In that case, the direction of propagation of the temperature signal $T_2(x,t)$ is the negative $x$ direction, namely the direction of increasing depth inside the region $x < 0$. In this direction, $T_2(x,t)$ is correctly damped in amplitude, as expected.

Figure~\ref{fig11} shows that both the solutions $T_1(x,t)$ and $T_2(x,t)$ are involved if we consider an infinite solid medium extending over $-\infty < x < +\infty$, with a localised source of harmonic temperature signals at $x=0$. In this case, $T_1(x,t)$ describes the signal propagation over the range $x>0$, while $T_2(x,t)$ describes the signal propagation over $x<0$. At every time $t$, the temperature distribution is an even function of $x$ defined in the range $-\infty < x < +\infty$.   

It must be pointed out that, for $T_1(x,t)$, the exponential growth rate, $s$, and the wavenumber $k$ turn out to be given by
\eqn{
s = -\sqrt{\frac{\omega}{2 \alpha}} \qc
k = \sqrt{\frac{\omega}{2 \alpha}}.
}{AA4}
On the other hand, for $T_2(x,t)$, the exponential growth rate and the wavenumber are given by 
\eqn{
s = \sqrt{\frac{\omega}{2 \alpha}} \qc
k = -\sqrt{\frac{\omega}{2 \alpha}}.
}{AA5}
Then, borrowing the definitions of spatial stability/instability given by \equasa{29ant1}{29ant2}, we can say that both solutions $T_1(x,t)$ and $T_2(x,t)$ are spatially stable $(s\, k < 0)$. 

\vfill

\end{document}